\documentclass{article}
\usepackage{spconf,amsmath,graphicx}
\usepackage{mathtools, cuted}
\usepackage{setspace}
\usepackage{flushend, cuted}
\usepackage{lipsum}
\usepackage{lingmacros}
\usepackage{tree-dvips}
\usepackage{varioref}
\usepackage{colortbl}
\usepackage{graphicx}
\usepackage{epstopdf}
\usepackage{color,soul}
\usepackage{amsfonts}
\usepackage{subfigure}
\usepackage{multicol}
\usepackage{multirow}
\usepackage{hhline}
\usepackage{amssymb}
\usepackage{newlfont}
\usepackage{mathrsfs}
\usepackage{cite}
\usepackage{nomencl}   
\usepackage{fancyhdr}
\usepackage{caption}

\usepackage{float}
\usepackage{placeins}
\usepackage{tabularx}
\usepackage[table]{xcolor}

\usepackage[linesnumbered,ruled]{algorithm2e}
\SetKwRepeat{Do}{do}{while}%

\DeclareRobustCommand{\frac}[3][0pt]{%
	{\begingroup\hspace{#1}#2\hspace{#1}\endgroup\over\hspace{#1}#3\hspace{#1}}}

\let\norm\undefined % <-- "Undefine" \norm
\DeclarePairedDelimiter\norm{\lVert}{\rVert}

\title{Analytic Optimization-Based Microbubble Tracking in \\ Ultrasound Super-Resolution Microscopy}
\name{Md Ashikuzzaman$^{a}$, Brandon Helfield$^{b, c}$, and Hassan Rivaz$^{a}$}
\address{$^{a}$ Department of Electrical and Computer Engineering, Concordia University, Montreal, QC, Canada\\
	$^{b}$ Department of Biology, Concordia University, Montreal, QC, Canada\\
	$^{c}$ Department of Physics, Concordia University, Montreal, QC, Canada
}
\begin{document}
	
	\setlength{\abovedisplayskip}{3pt}
	\setlength{\belowdisplayskip}{3pt}
	%\ninept
	%
	\maketitle
\begin{abstract}
\textbf{Ultrasound localization microscopy (ULM) refers to a promising medical imaging modality that systematically leverages the advantages of contrast-enhanced ultrasound (CEUS) to surpass the diffraction barrier and delineate the microvascular map. Localization and tracking of microbubbles (MBs), two significant steps of ULM, facilitate generating the vascular map and the velocity distribution, respectively. Herein, we propose a novel MB tracking technique considering temporal pairing as a bubble-set registration problem. Iterative registration is performed between the bubble sets in two consecutive time instants by analytically optimizing a cost function that takes position and point-spread function (PSF) similarities as well as physically plausible levels of bubbles' movement into account. Furthermore, we infer MBs' parity in a fuzzy manner instead of binary. The proposed technique performs well in validation experiments with two synthetic and two \textit{in vivo} datasets provided by the Ultrasound Localisation and TRacking Algorithms for Super Resolution (ULTRA-SR) Challenge.}                            
\end{abstract}


\begin{figure}[h]
	%\DeclareGraphicsExtensions{.eps}
	\centering
	\subfigure[CEUS image with bubble centers]{{\includegraphics[width=.165\textwidth]{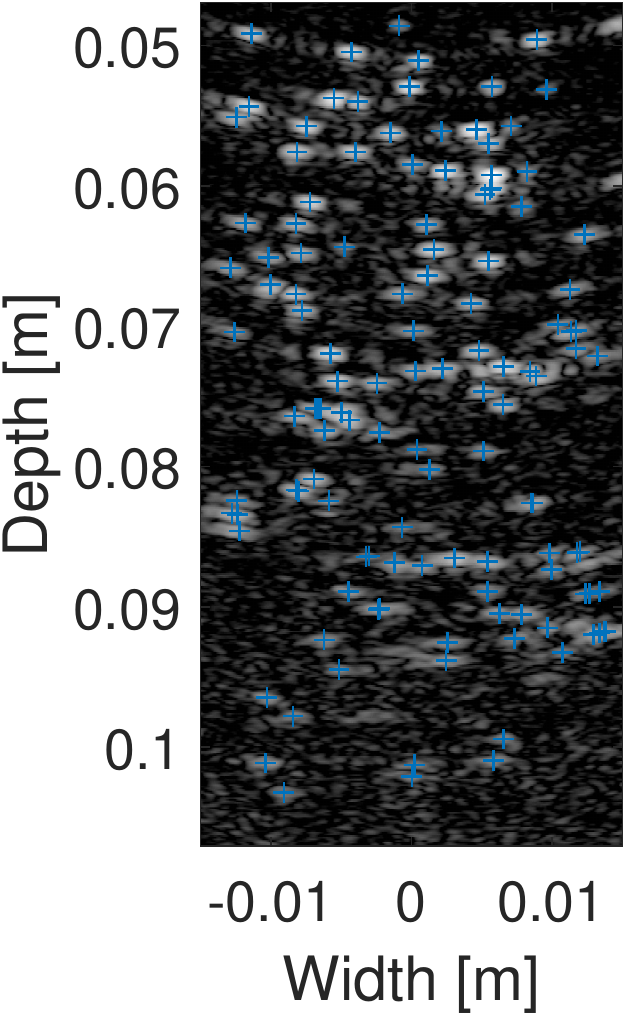}}}%
	\subfigure[Bubble density]{{\includegraphics[width=.165\textwidth]{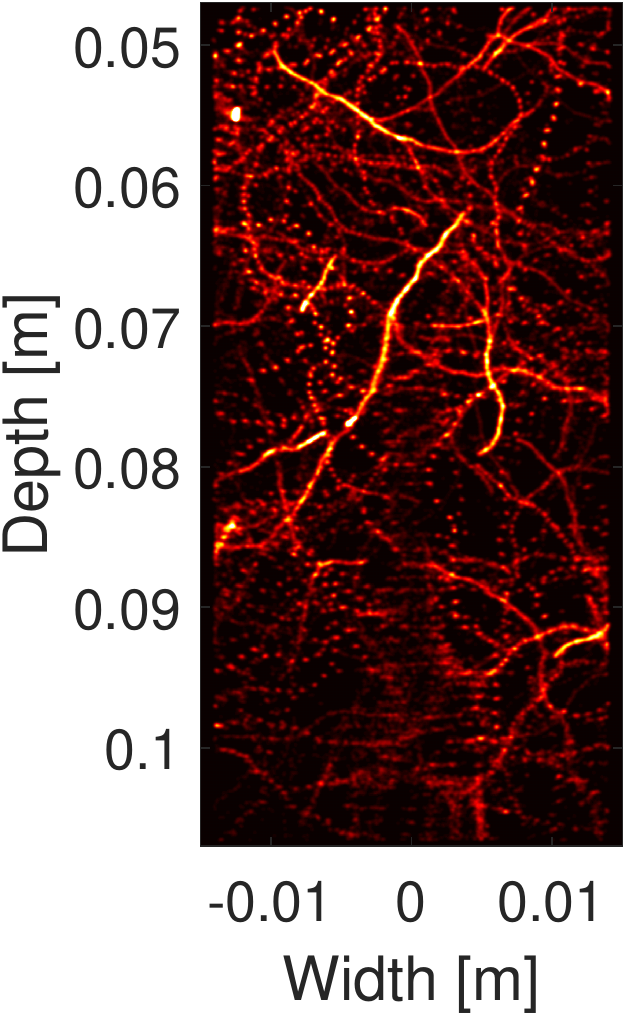}}}%
	\subfigure[Speed]{{\includegraphics[width=.165\textwidth]{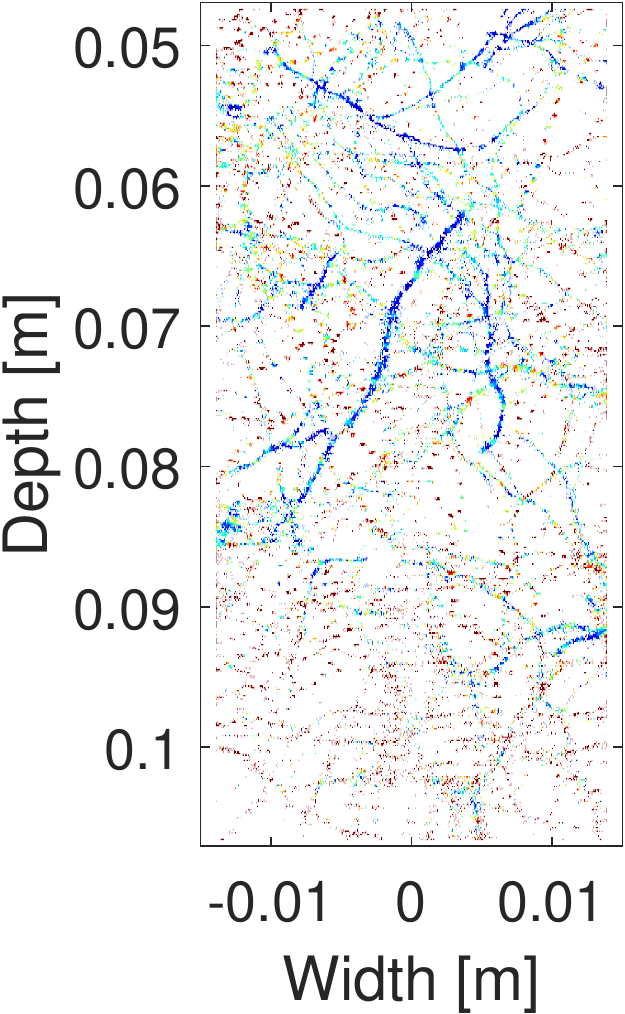}}}
	\subfigure[Color bar, density]{{\includegraphics[width=.23\textwidth]{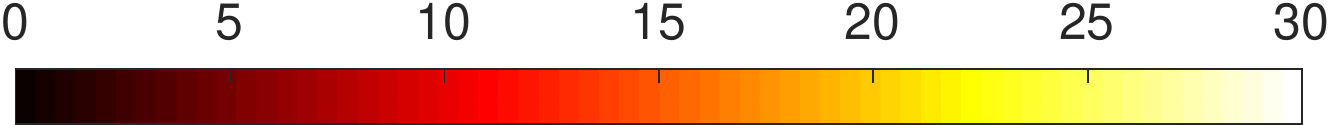}}}%
	\quad
	\subfigure[Color bar, speed ($ms^{-1}$)]{{\includegraphics[width=.23\textwidth]{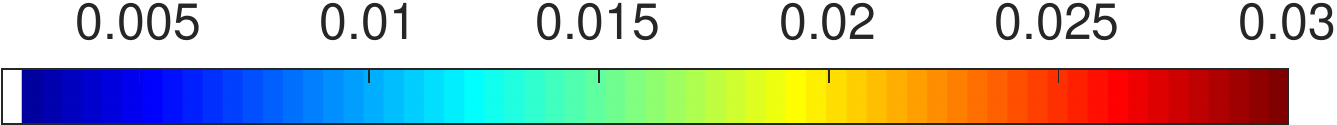}}}%
	\caption{Results for the low-frequency synthetic dataset. (a) shows a CEUS image with the localized bubble centers, whereas (b) and (c) depict the localization density and speed maps, respectively.}
	\label{simu1}
\end{figure}

\section{Introduction}
\label{sec:intro}
Inspired by its optical analog, ultrasound localization microscopy (ULM)~\cite{christensen2014vivo} has evolved to break conventional ultrasound's diffraction barrier and provide a super-resolution (SR) vascular map. ULM is a minimally-invasive imaging technique that processes contrast-enhanced ultrasound (CEUS) image sequences to reveal the microvascular network of interest. SR ultrasound research has seen rapid advancement during the last decade. Thus, ULM has established itself as a promising modality to assist the assessment of vascular health~\cite{demene2021transcranial,demeulenaere2022coronary} and the diagnoses of fatal diseases~\cite{hao2019non}.

The localization and tracking of gas-filled microbubbles (MBs) circulating through the bloodstream are two vital steps of ULM. Detecting the MBs and localizing their centers facilitates generating the localization density map representing the vascular network under investigation. Among many MB localization techniques available in the literature, the thresholding and amplitude-weighted coordinate averaging-based approach~\cite{christensen2014vivo} is a mainstream one. Along this line, a pipeline technique involving rough MB center localization based on the correlation map between the CEUS image and the point-spread function (PSF) followed by a center of mass calculation step has been proposed~\cite{tang2020kalman}. Sparsity-based deconvolution was proposed in \cite{yan2022super} to improve the isolation of overlapping bubbles. In \cite{hardy2021sparse}, the probe channels have been randomly subsampled to reduce ULM's computational load and hardware complexity, minimally affecting the MB localization accuracy. Deep neural network (DNN)-based techniques~\cite{brown2021faster} have also been proposed for improved MB tracking.       

Tracking the localized MBs is required to produce the velocity map. MB tracking is predominantly accomplished by block-matching~\cite{christensen2014vivo} or bipartite pairing-based~\cite{song2017improved} techniques. A Kalman filter-based robust microbubble tracking technique has been developed in \cite{tang2020kalman}. Another robust technique employing an echo-Lagrangian particle filter has been proposed in \cite{jeronimo2020echo}. Hansen \textit{et al.}~\cite{hansen2016robust} have devised a cross-correlation-based motion compensation technique for improved MB tracking. A Euclidian distance-based nearest neighbor algorithm is exploited in \cite{johnson2020three}, whereas a forward-backward strategy is adopted in \cite{taghavi2022microbubble} for tracking the MBs in ULM.     

For localizing the MBs, we use a standard correlation- and amplitude-weighted center of mass-based technique. For MB tracking, we have developed an analytic optimization-based technique inspired by its success in several research fields~\cite{guest,soul,soulmate,rglue_tuffc,chui2003new}. We look at the temporal MB tracking problem from the standpoint of bubble-set registration and make bubble-pairing decisions probabilistically. The reference and target bubble sets are registered through iterative optimization of a cost function penalizing position dissimilarity, PSF disparity, and large MB movement. The proposed ULM technique is validated against low- and high-frequency synthetic datasets, \textit{in vivo} rat brain, and human lymph node datasets.                                                            

\begin{figure}
	%\DeclareGraphicsExtensions{.eps}
	\centering
	\subfigure[CEUS image with bubble centers]{{\includegraphics[width=.165\textwidth]{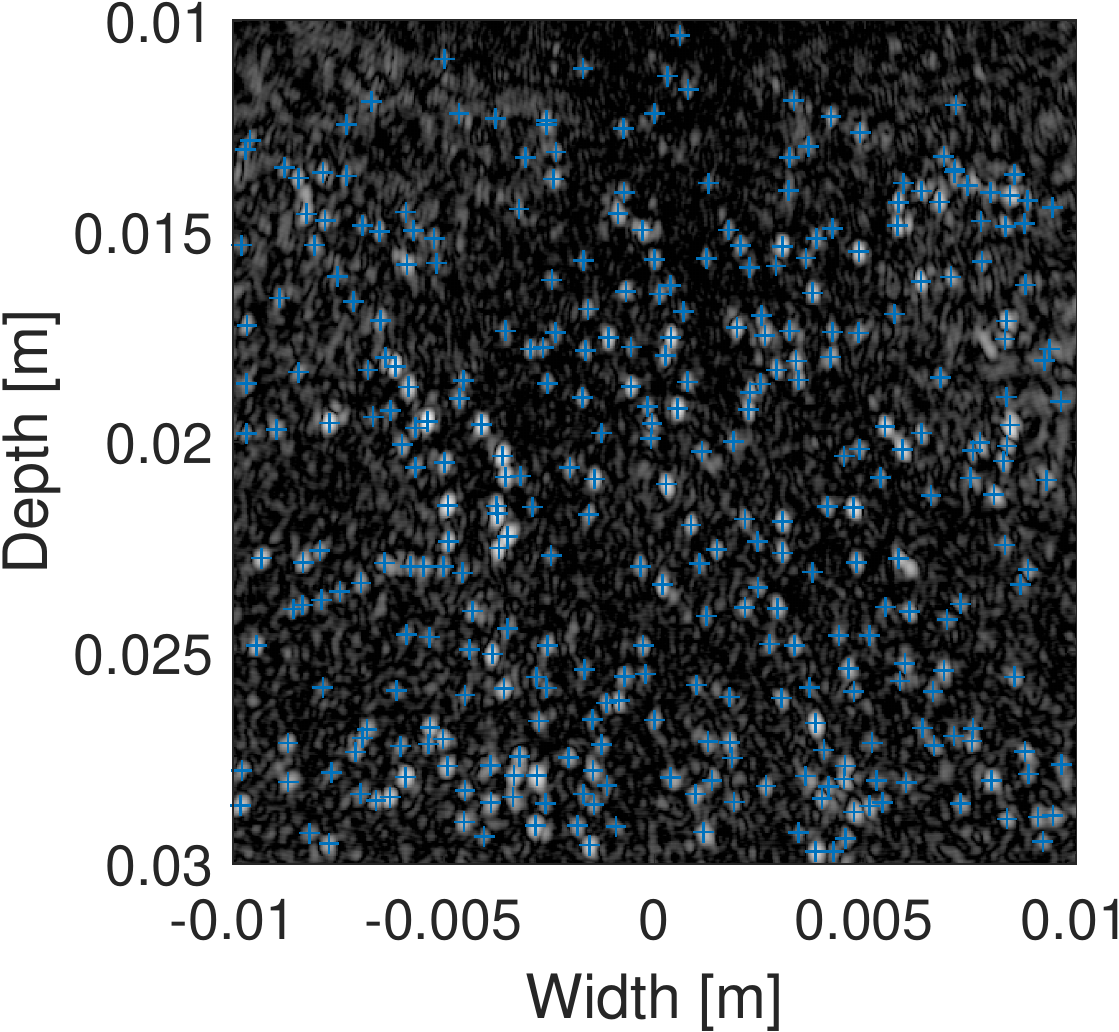}}}%
	\subfigure[Bubble density]{{\includegraphics[width=.165\textwidth]{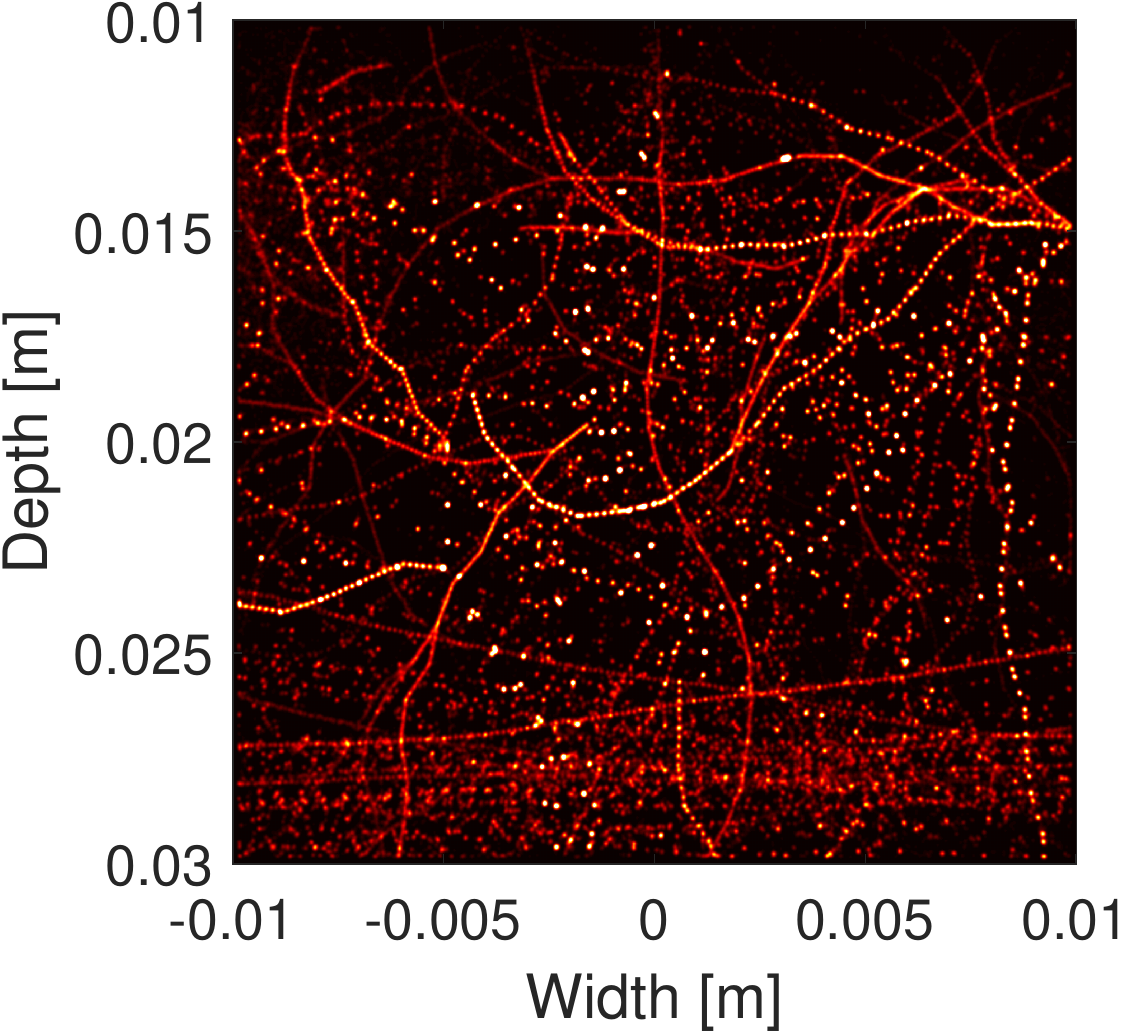}}}%
	\subfigure[Speed]{{\includegraphics[width=.165\textwidth]{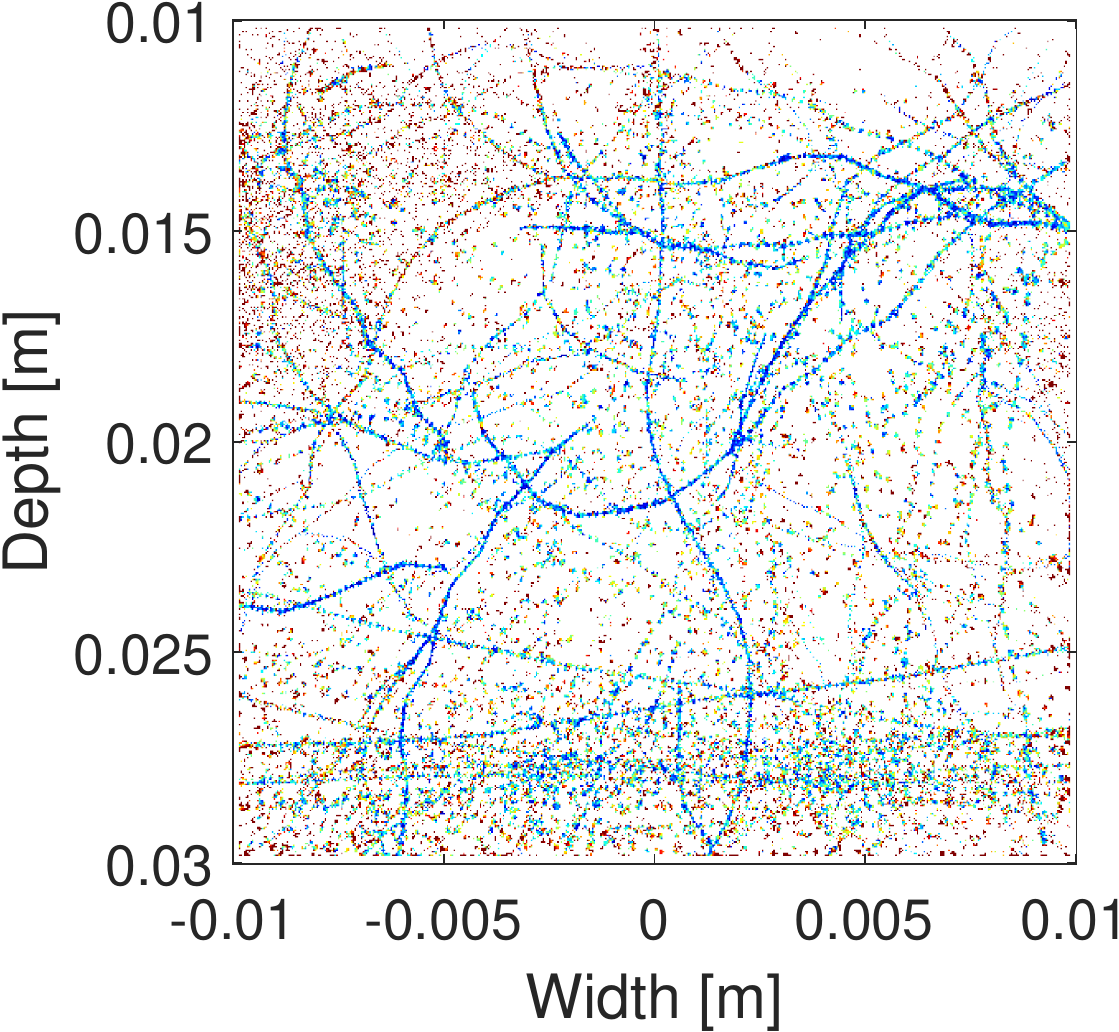}}}
	\subfigure[Color bar, density]{{\includegraphics[width=.23\textwidth]{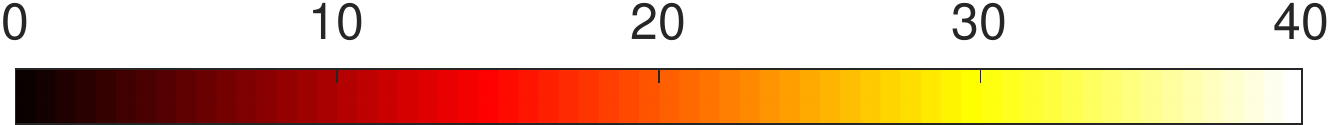}}}%
	\quad
	\subfigure[Color bar, speed ($ms^{-1}$)]{{\includegraphics[width=.23\textwidth]{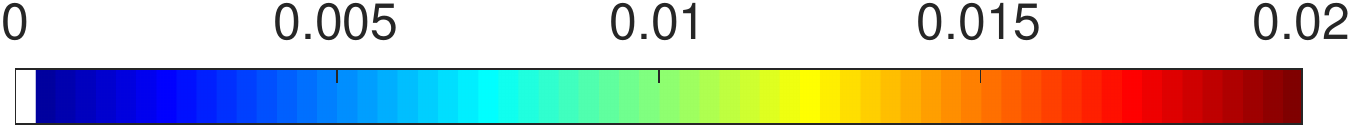}}}%
	\caption{Results for the high-frequency synthetic dataset. (a) presents a CEUS image with the localized bubble centers, whereas (b) and (c) show the localization density and speed maps, respectively.}
	\label{simu2}
\end{figure}

\begin{figure}
	%\DeclareGraphicsExtensions{.eps}
	\centering
	\subfigure[CEUS image with bubble centers]{{\includegraphics[width=.165\textwidth]{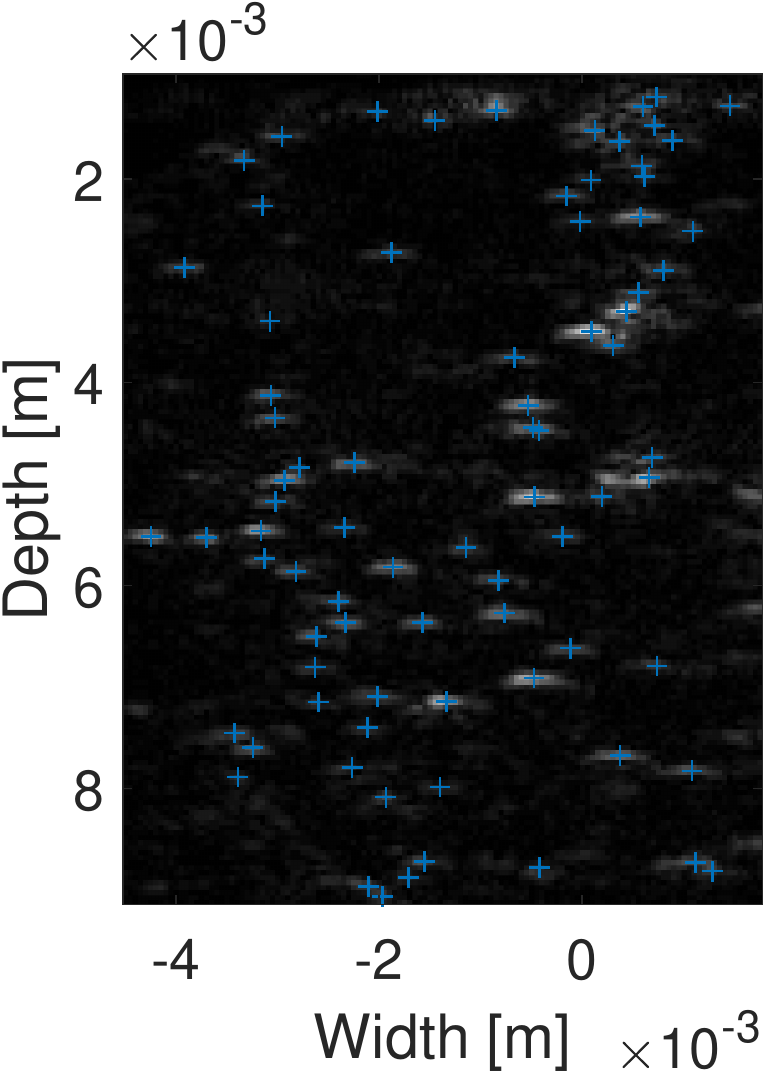}}}%
	\subfigure[Bubble density]{{\includegraphics[width=.165\textwidth]{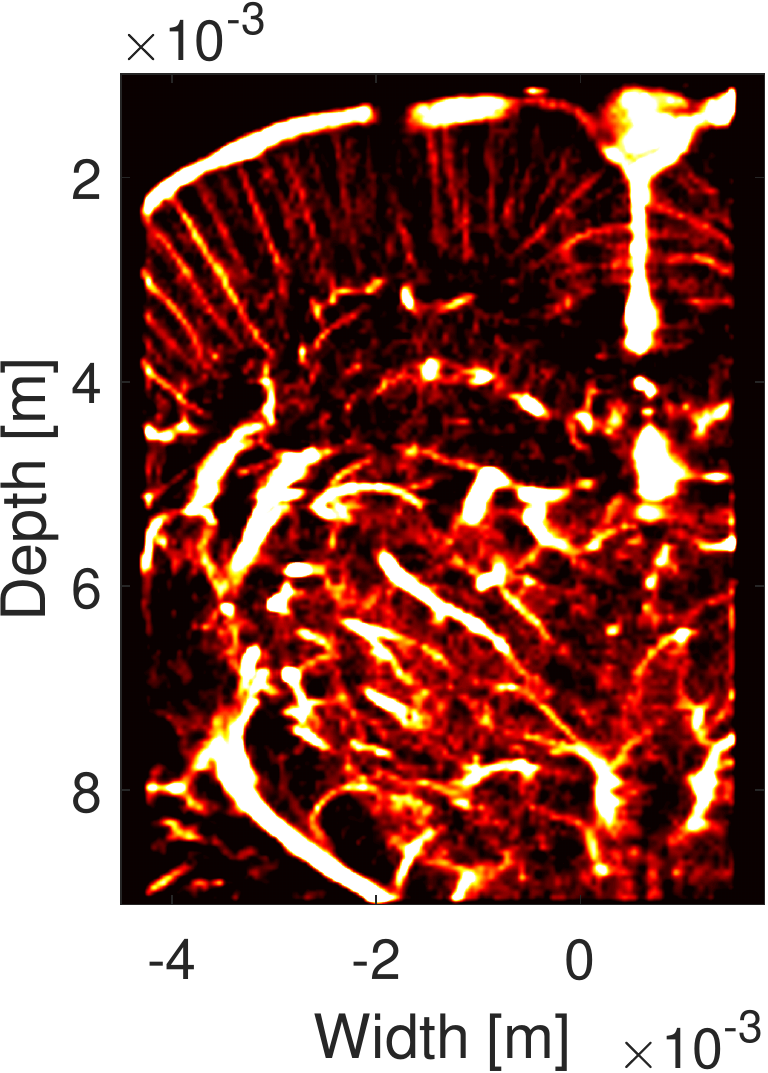}}}%
	\subfigure[Speed]{{\includegraphics[width=.165\textwidth]{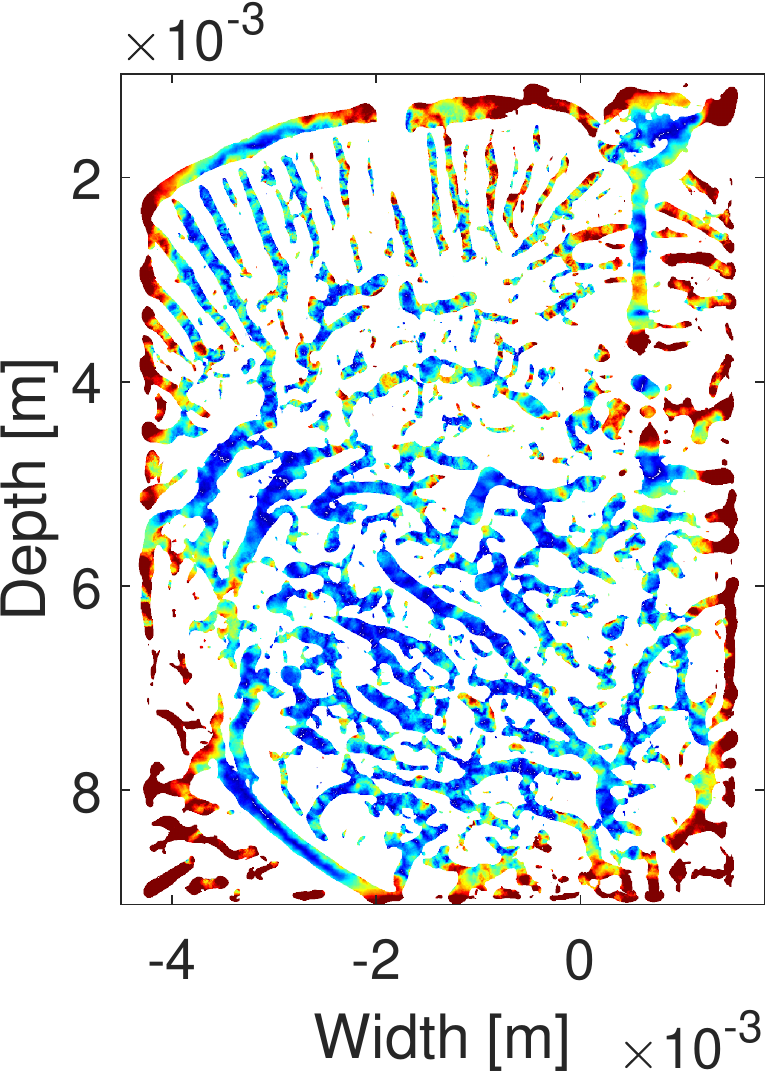}}}
	\subfigure[Color bar, density]{{\includegraphics[width=.23\textwidth]{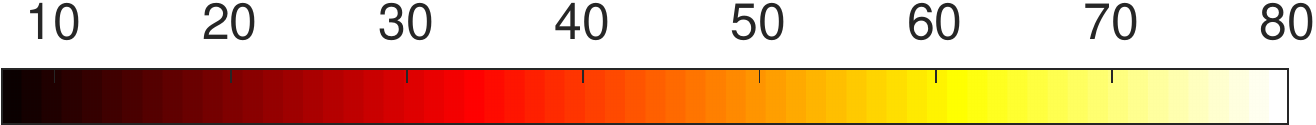}}}%
	\quad
	\subfigure[Color bar, speed ($ms^{-1}$)]{{\includegraphics[width=.23\textwidth]{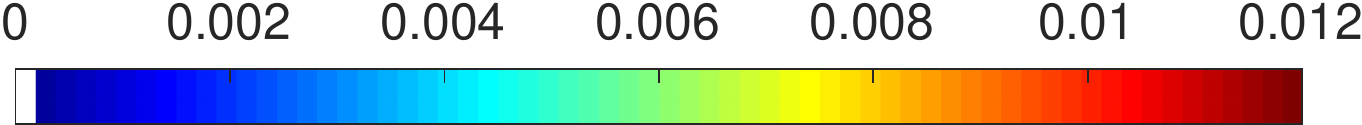}}}%
	\caption{Results for the \textit{in vivo} rat brain dataset. (a) shows a CEUS image with the localized MB centers, whereas (b) and (c) depict the bubble density and speed maps, respectively.}
	\label{brain}
\end{figure}

\section{Methods}
\label{sec:method}
In this section, we first describe our MB localization and the bubble density map rendering technique. Then we outline the proposed MB tracking algorithm and its mathematical details.  

\subsection{MB Localization and Bubble Density Map}
A suitable bubble signal was extracted from one of the CEUS images and considered the PSF. The local peaks of the correlations between the PSF and the CEUS image patches determined the bubble centers' integer locations. The subpixel localization precision was achieved by amplitude-weighted averaging of the grid coordinates around the integer center location. Finally, a Gaussian around each localized MB center was considered, and all such Gaussians were added to produce the SR bubble density map.

\begin{figure*}[h]
	%\DeclareGraphicsExtensions{.eps}
	\centering
	\subfigure[CEUS image with bubble centers]{{\includegraphics[width=.28\textwidth]{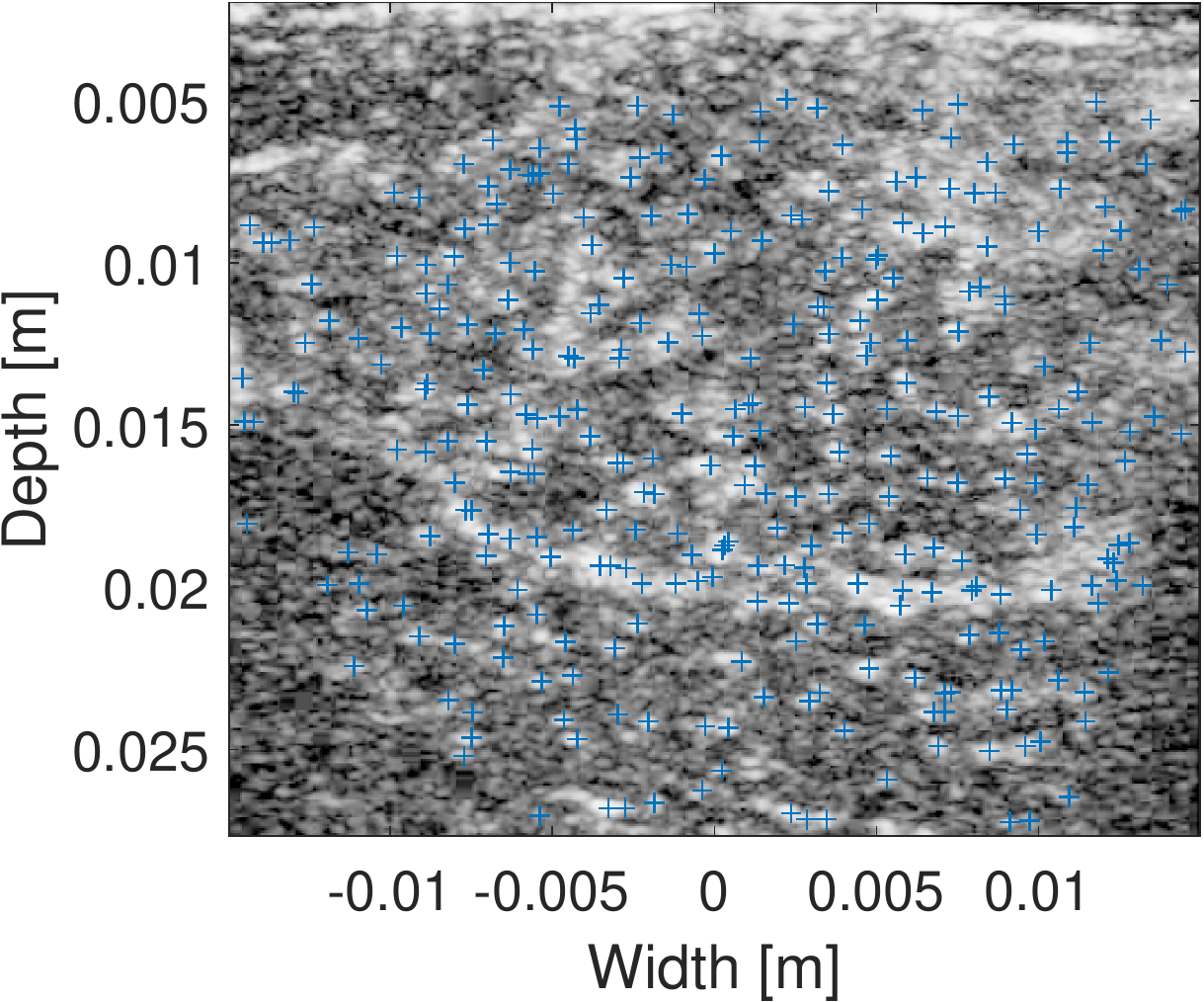}}}%
	\subfigure[Bubble density]{{\includegraphics[width=.28\textwidth]{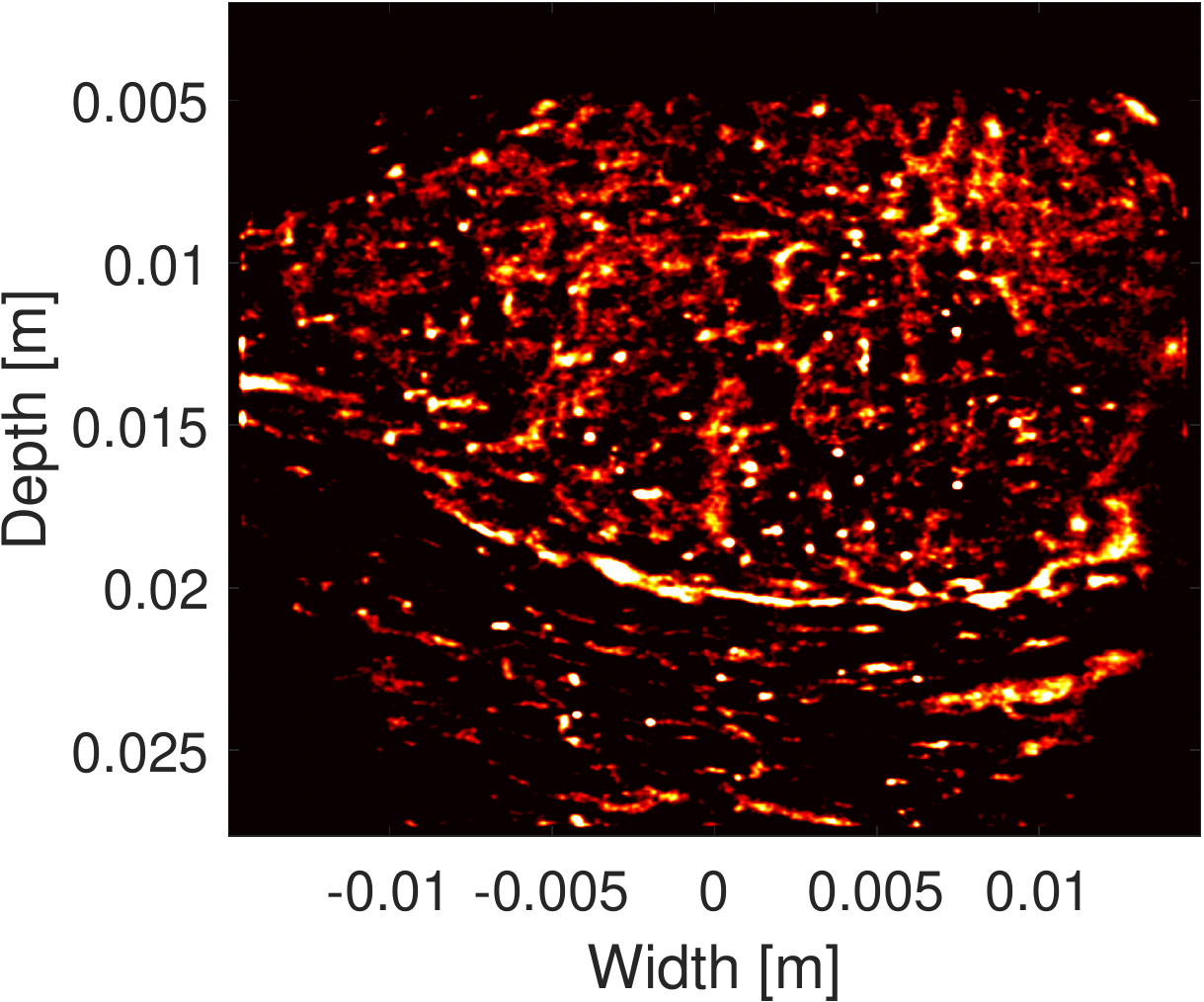}}}%
	\subfigure[Speed ($ms^{-1}$)]{{\includegraphics[width=.28\textwidth]{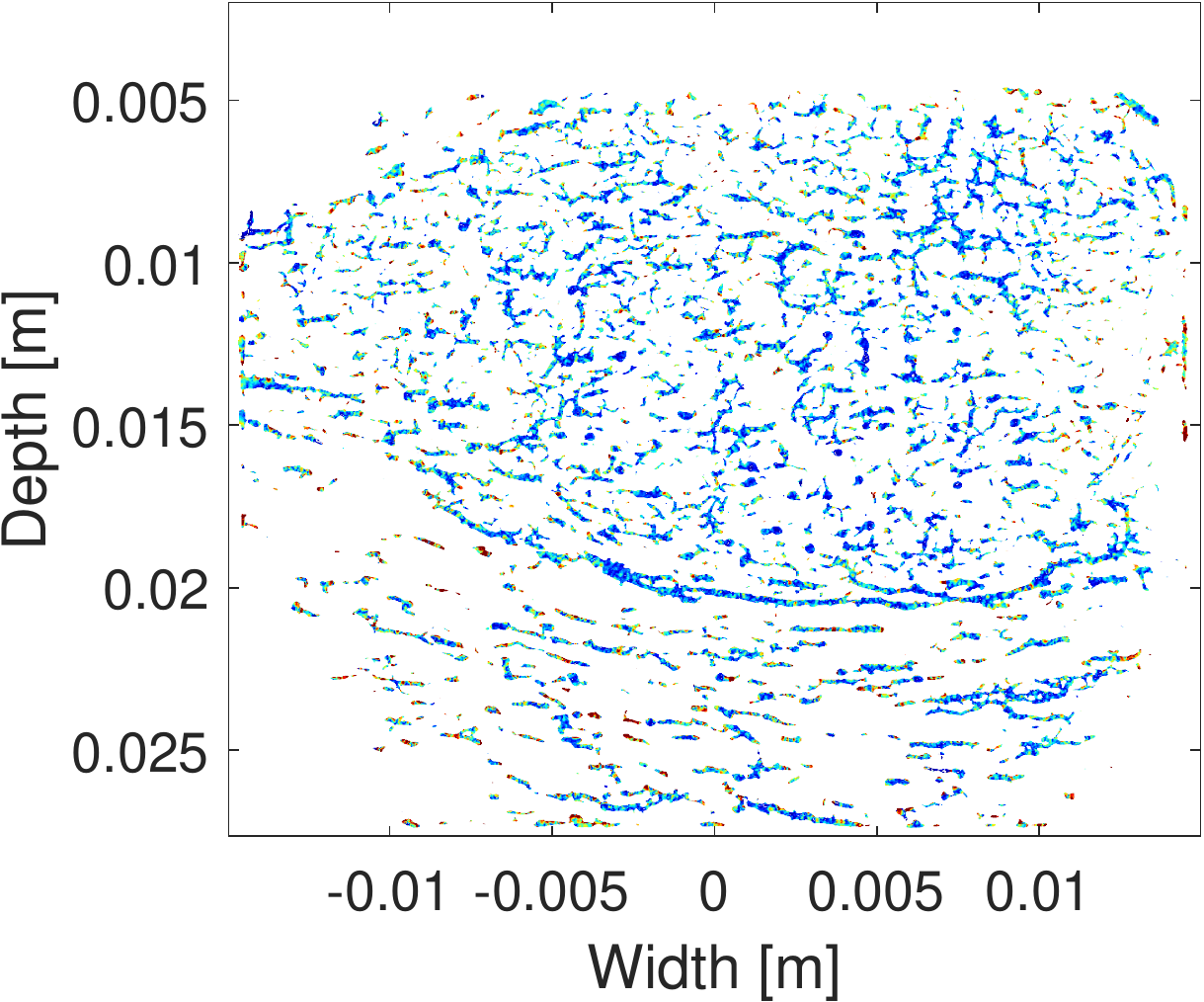}}}%
	\quad
	\subfigure[]{{\includegraphics[height=0.233\textwidth]{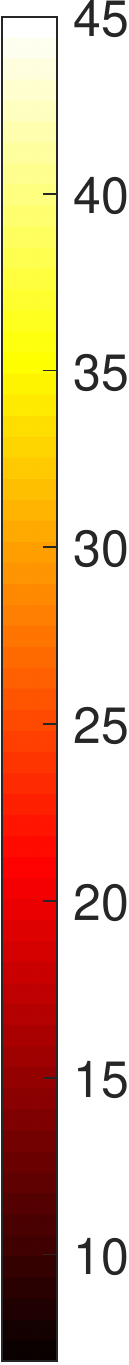}}}%
	\quad
	\subfigure[]{{\includegraphics[height=0.233\textwidth]{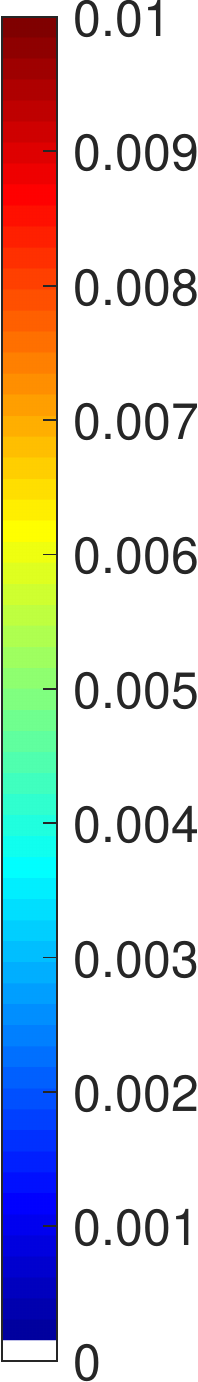}}}%
	\caption{Results for the lymph node dataset. (a) manifests a CEUS image with the localized MB centers, whereas (b) and (c) present the MB density and speed maps, respectively. (d) and (e) show the density and speed color bars, respectively.}
	\label{lymph}
\end{figure*}

\subsection{MB Tracking and Speed Map}
Let $r_{i}$, $i \in \{1,2,3,...,m\}$ and $t_{j}$, $j \in \{1,2,3,...,n\}$ denote the reference and target bubble sets, respectively. $x_{i}$ and $y_{j}$ refer to the reference and target MB center locations, respectively. We aim to determine the correspondence between the reference and target MBs by obtaining the transformation function $f$ that registers the two bubble sets. To that end, we alternatingly iterate between optimizing a cost function $C$ to obtain $f$ and updating the bubble-pairing probability $p_{i,j}$. As given in Eq.~\ref{eq:c}, $C$ penalizes position and PSF disparities, and forces the estimated bubble movements to be small.    

\begin{equation}
\begin{aligned}
&C = \alpha \sum\limits_{j=1}^n \sum\limits_{i=1}^m p_{i,j} \norm{x_{i}-f(y_{j})}^{2} +\\
&\beta \sum\limits_{j=1}^n \sum\limits_{i=1}^m p_{i,j} \norm{PSF_{x_{i}}-PSF_{f(y_{j})}}^{2} + \gamma \sum \norm{f}^{2}
\end{aligned}
\label{eq:c}
\end{equation}

\noindent
where $\alpha$, $\beta$, and $\gamma$ control the relative importance of the constraints. $C$ is analytically minimized using the Gauss-Newton optimization technique. In each iteration, $p_{i,j}$ is updated by combining the location and PSF similarity-based matching probabilities: 

\begin{equation}
\begin{aligned}
p_{i,j} = p_{i,j,loc} \times p_{i,j,psf}
\end{aligned}
\end{equation}

\noindent
where $p_{i,j,loc}$ and $p_{i,j,psf}$ are defined as follows. 

\begin{equation}
\begin{aligned}
p_{i,j,loc} = \frac{1}{w_{1}} e^{-\frac{[d(x_{i},f(y_{j}))]^{2}}{2w_{1}}}
\end{aligned}
\end{equation}

\begin{equation}
\begin{aligned}
p_{i,j,psf} = \frac{1}{w_{2}} e^{-\frac{\sum [\textrm{$PSF$ disparity}]^{2}}{2w_{2}}}
\end{aligned}
\end{equation}

\noindent
Here, $w_{1}$ and $w_{2}$ are tunable weights, whereas $d(\cdot)$ stands for the Euclidian distance. Row and columns of the $m \times n$-sized probability matrix are iteratively normalized so that the entries of each row and column sum up to one. A reference bubble is finally paired to one of the target bubbles based on the highest matching probability.

The velocity of each MB is estimated by frame-to-frame tracking. The velocities of all MBs localized within a circle centered at a particular grid point are gathered to form the final velocity map. A principal component analysis (PCA) is performed to reject the velocities more than three standard deviations away from the two orthogonal axes. After this outlier rejection step, a distance-based Gaussian-weighted averaging is performed on the remaining velocities to find the final velocity at the grid location. This pipeline provides the velocity at each grid point, and the amplitudes of the $2D$ velocities render the speed map.   

%\vspace{0.05in}
\section{Results and Discussion}
We have validated the proposed ULM technique against low- and high-frequency synthetic as well as \textit{in vivo} rat brain and human lymph node datasets provided by the ULTRA-SR Challenge. The low- and high-frequency datasets were simulated with a GE M5Sc-D phased array and a Verasonics L11-4v linear array transducer setting the center frequencies to 2.84 MHz and 7.24 MHz, respectively. The brain and lymph node datasets, respectively, were acquired at 15.63 MHz and 5.6 MHz transmit frequencies. The number of CEUS images used for generating the SR images corresponding to the low- and high-frequency synthetic and \textit{in vivo} rat and human datasets were 500, 500, 8000, and 1382, respectively.

\subsection{Synthetic Datasets}
Figs.~\ref{simu1} and \ref{simu2} show representative CEUS images with localized bubble centers, SR localization density maps, and speed images for the low- and high-frequency synthetic datasets, respectively. The localized MB centers and the bubble density maps validate the merit of the localization technique employed in this work. The speed maps clearly distinguish between the high- and low-speed simulated vessels. In addition, the speed distributions correspond well with the CEUS videos' visual assessment. The solid and dotted vessels on the localization density images appear as low- and high-speed vessels, respectively, on the speed maps, which further demonstrate the proposed tracking technique's potential.

\subsection{\textit{In Vivo} Rat Brain Dataset}
The \textit{in vivo} rat brain results have been shown in Fig.~\ref{brain}. It is evident from the localized centers that almost all visible MBs are detected. The localization density map visualizes the microvasculature and the relatively large vessels. Note that the vascular network delineated by our technique is consistent with the previously published rat brain vasculature~\cite{hingot2019microvascular}. The speed map yields good contrast between the high- and low-speed vascular regions. In addition, the speed distribution agrees well with the visual judgment of the MBs' speed in different spatial regions of the CEUS sequences.

\subsection{\textit{In Vivo} Human Lymph Node Dataset}
Fig.~\ref{lymph} presents a CEUS image with MB centers and the SR density and speed maps for the human lymph node dataset. Given the difficulty of this particular dataset, the localization technique exhibits good performance, which is evident in the localization density map. The speed map distinguishes the low-speed vessels from the high-speed ones and shows good correspondence with our visual perception of the comparative speed distribution in the CEUS videos.  

\section{Conclusion}      
A ULM pipeline is proposed where MB is localized using a cross-correlation-based technique. We have devised a novel MB tracking algorithm that registers the reference bubble set to the target one by iteratively optimizing a cost function. Instead of binary, MB pairing is modeled as a fuzzy system where bubbles are temporally matched through probabilistic inferences. The proposed technique performs well on synthetic and \textit{in vivo} datasets.   	
	
\section*{\normalsize{ACKNOWLEDGMENT}}    
This work is partly supported by the Natural Sciences and Engineering Research Council of Canada (NSERC). Md Ashikuzzaman holds the B2X Doctoral Research Fellowship provided by the Fonds de Recherche du Québec - Nature et Technologies (FRQNT). We thank the organizers of the ULTRA-SR Challenge for providing the synthetic and \textit{in vivo} datasets.

	\FloatBarrier
	\bibliographystyle{IEEEtran}
	\bibliography{ref}
	
\end{document}